\documentclass[letterpaper,twocolumn,10pt]{article}
\usepackage{usenix}

\usepackage{tikz}
\usepackage{amsmath}

\usepackage{filecontents}

\usepackage{etoolbox}
\usepackage{booktabs}
\usepackage{hyperref}
\usepackage{tikz}
\usepackage{amsmath}
\usepackage{threeparttable}
\usepackage{filecontents}
\usepackage{graphicx}
\usepackage{fancyvrb}
\usepackage{listings}
\usepackage{caption}
\usepackage{framed}
\usepackage{xspace}
\usepackage{url}
\usepackage{setspace}
\usepackage{multirow}
\usepackage{newfloat}
\DeclareFloatingEnvironment[fileext=lol, listname={List of Listings}, name=Listing, placement=tbhp, within=none]{listing}

\lstdefinestyle{aijonBase}{
    basicstyle=\ttfamily\scriptsize,
    numbers=left,
    numberstyle=\tiny,
    numbersep=5pt,
    xleftmargin=1.0em,
    xrightmargin=0pt,
    breaklines=true,
    breakatwhitespace=false,
    keepspaces=true,
    columns=fullflexible,
    showstringspaces=false,
    frame=none,
    tabsize=4,
    upquote=true,
}
\usepackage{dblfloatfix}
\usepackage{enumitem}
\usepackage{adjustbox}
\usepackage{pifont}
\usepackage{tcolorbox}

\hypersetup{
    colorlinks,
    linkcolor={red!80!black},
    citecolor={blue},
    urlcolor={blue!80!black}
}

\definecolor{maroon}{rgb}{0.5,0,0}

\definecolor{pinegreen}{rgb}{0.0, 0.47, 0.44}
\definecolor{rossocorsa}{rgb}{0.83, 0.0, 0.0}

\newcommand{\cmark}{\textcolor{pinegreen}{\ding{51}}}
\newcommand{\xmark}{\textcolor{rossocorsa}{\ding{55}}}

\definecolor{tick-green}{RGB}{153,255,153}
\definecolor{cross-red}{RGB}{255,153,153}
\definecolor{yellow}{rgb}{0.99, 0.93, 0}

\newcommand{\aijon}{\textsc{AIJON}\xspace}

\newcommand{\reachedmagmabugs}{72\xspace}
\newcommand{\aflppmagmatriggered}{38\xspace}
\newcommand{\ijonmagmatriggered}{37\xspace}
\newcommand{\aijonmagmaannotated}{71\xspace}
\newcommand{\aijonmagmatriggered}{34\xspace}
\newcommand{\aijonmissed}{4\xspace}
\newcommand{\aijonfaster}{16\xspace}

\newcommand{\aflppfasteraijon}{18\xspace}
\newcommand{\ijonaflppcommon}{35\xspace}
\newcommand{\ijonfasteraflpp}{14\xspace}

\begin{document}

\date{}

\title{\Large \bf AIJon: Automated Generation of Annotations for Fuzzing}

\author{
{\rm Jayakrishna Menon Vadayath$^{*}$, Hulin Wang$^{*}$, Moritz Schloegel$^{\dagger}$, Jie Hu$^{*}$, Wil Gibbs$^{*}$,}\\
{\rm Tiffany Bao$^{*}$, Adam Doup\'e$^{*}$, Ruoyu ``Fish'' Wang$^{*}$, Yan Shoshitaishvili$^{*}$}\\[4pt]
$^{*}$Arizona State University, \quad $^{\dagger}$CISPA Helmholtz Center for Information Security\\
$^{*}$\textit{\{jvadayat,hwang551,jiehu12,wfgibbs,tbao,doupe,fishw,yans\}@asu.edu} \quad $^{\dagger}$\textit{moritz.schloegel@cispa.de}
} 

\maketitle

\begin{abstract}
Modern fuzzers use code coverage as feedback to guide their exploration which has proven to be an effective strategy for driving exploration. However, this strategy overlooks inputs that may be interesting to the target program even without uncovering new code paths.
Fortunately, prior research has shown that annotations generated by human domain experts can provide additional feedback, guiding the fuzzer towards interesting parts of the program.

In this paper, we replicate experiments presented in IJON and extend them to real-world vulnerability detection at scale.
To mitigate the scalability challenge, imposed by the need for human domain expertise, we propose utilizing LLMs to automatically generate annotations.
We demonstrate the applicability of LLMs for this purpose and observe that LLMs can generate annotations that perform comparably to human-generated annotations.

Motivated by this finding, we design AIJON, a system that leverages LLMs to automatically generate IJON-style annotations.
We evaluate AIJON on the Magma benchmark and surprisingly observe that annotation-based fuzzing does not perform strictly better than AFL++.
We conduct several experiments to identify the cause of our results and identify key insights regarding the impact of annotations on fuzzing campaigns, including their effect on the energy distribution of the fuzzer.
Notably, we observe that LLMs can generate annotations that achieve comparable results to human generated ones, thus opening the door for future research to perform further studies on the impact of annotations at scale.

\end{abstract}

%

\section{Introduction}

Fuzz testing, or fuzzing, has proven one of the most successful approaches to discovering vulnerabilities in software, featuring low to zero false positives while requiring minimal human effort.
Today, large codebases such as the Linux kernel or Chromium employ fuzzing to find bugs in their code before users are impacted.
%
Modern fuzzers, such as AFLplusplus~\cite{fioraldi2020aflpp}, HonggFuzz~\cite{honggfuzz}, LibFuzzer~\cite{libfuzzer}, or Syzkaller~\cite{syzkaller}, combine high throughput with a clever heuristic for steering exploration: code coverage. 
Usually, by instrumenting the program at compile time, the fuzzer receives feedback on every executed input. 
This allows it to keep in its queue only inputs that uncover novel program behavior, for example, a new edge in the control-flow graph. 
In other words, the fuzzers' attention is continuously steered towards exercising unseen program behavior.
Despite being highly efficient at driving exploration, this strategy overlooks that inputs can be interesting \emph{even when they do not uncover new program behavior}~\cite{schiller2025novelty,fioraldi2021use}. 
For programs with complex state machines, driving exploration toward unseen behavior may simply not be sufficient to explore a program \emph{effectively}.
Aschermann et al.~\cite{aschermann2020ijon} showed this convincingly for specific programs that were considered out-of-reach for traditional fuzzers, such as \textit{Super Mario Bros.} or the infamous Maze.
Their proposed solution is to ask a human domain expert to insert \emph{annotations} into the code that provide the fuzzer with \emph{additional} feedback, helping it explore interesting parts of the program.
Essentially, this approach harnesses human expertise to identify relevant patterns or states that are worthwhile for the fuzzer to pursue.
The initial experiments looked very promising, and the fuzzing community has widely regarded these IJON annotations as an effective mechanism of helping the fuzzer using human insight. Yet, to date this requirement of human expertise has lead to no large-scale study of such annotations, leaving their usefulness for a fuzzer's true goal, finding bugs, in the dark.

In this paper, we strive to replicate the experiments presented in IJON and, crucially, extend them to real-world vulnerability detection at scale.
Such scalability is only possible when we remove IJON's need for a human domain expert.
Fortunately, recent advances in Large Language Models (LLMs) have significantly improved their ability to comprehend and generate syntactically correct code.
Given the powerful capabilities of LLMs, we propose using LLMs to automatically generate IJON-style annotations for target programs, thereby eliminating the need for human expertise.
In a first step, we validate the applicability of LLMs for generating IJON-style annotations; to this end, we replicate the \textit{Super Mario Bros.} experiment of IJON, but we query an LLM to generate annotations for the game. 
We then evaluate the impact of these LLM-generated annotations on their ability to guide the fuzzer toward completing levels in the game and compare the results with those obtained by IJON.
We observe that the LLM can indeed generate annotations that perform comparably to those generated by the authors of IJON.

Motivated by this finding, we now attempt what IJON could not do for the lack of automation: We explore the applicability of LLMs in generating annotations for vulnerability detection in real-world software at scale.
To perform such an analysis, we first develop \aijon, a system that leverages LLMs to automatically generate IJON-style annotations for target programs.
Using \aijon, we then conduct large-scale evaluations of annotations on the Magma benchmark~\cite{Hazimeh:2020:Magma}, a widely accepted bug benchmark with ground truth data.
Our experiments show counterintuitive results: we did not observe annotation-based fuzzing performing strictly better than the AFL++ baseline. 
While it triggers bugs faster in \aijonfaster cases, it slows down the fuzzer for \aflppfasteraijon vulnerabilities. 
Our first intuition may be to blame this on LLMs or our LLM-based implementation.
To test this hypothesis, we conduct two further experiments demonstrating that the LLM performs as expected and that human domain expert-generated annotations perform similarly.

A second hypothesis could be wrong use of annotations: In our evaluation on the Magma dataset, we inserted annotations for multiple vulnerabilities into the source code of the target programs.
To understand whether the presence of multiple annotated vulnerabilities was leading to sub-optimal performance, we conducted an additional experiment in which we generated a copy of each target program with annotations for only a single vulnerability.
We repeated this experiment for annotations generated by both \aijon and human-generated ones.
Our results indicate that even when only a single vulnerability is annotated, the performance of the fuzzer does not improve significantly across all vulnerabilities.

This led us to question the \emph{fuzzer's energy distribution} and to measure it using the fine-grained metrics provided by the Magma benchmark.
Naturally, annotations lead to additional inputs being saved to the queue, shifting the fuzzer's attention towards specific code regions.
Yet, such an attention shift does not always align with the location or path to a vulnerability.
For example, vulnerability TIF012 was reached only 4 million times before it was triggered, but by the end of 24 hours of fuzzing with IJON annotations, it had been reached almost 64 million times---a 16-fold increase attributable to targeted feedback from annotations.
Notably, most of this energy was spent after the vulnerability was discovered. 
As a result, annotations in TIF012 guided the fuzzer to overspend energy on program states with no additional vulnerabilities.  
We observe that annotations can lead to a significant shift in the fuzzer's energy distribution, which affects the overall fuzzing efficiency.
Moreover, we demonstrate that even when annotating a single vulnerability at a time, the presence of annotations may not always lead to improved performance.
Finally, we demonstrate that LLM-generated annotations perform comparably to human-generated annotations. This opens the door for future research to study the impact of annotations at scale without relying on human domain experts to generate annotations.

\textbf{Contributions.} In summary, our contributions are:
\begin{itemize}[noitemsep, topsep=0pt]
    \item We design a scalable approach to replicate IJON and apply it to real-world programs. 
    \item Based on our LLM-based annotation design, we develop \aijon, an automatic annotation system that does not require human domain experts and is compatible with AFL++.
    \item We conduct the first large-scale evaluation of annotations in fuzzing by comparing \aijon against a baseline AFL++ configuration without annotations on the Magma benchmark suite, studying the impact of annotations on vulnerability survival times.
    \item We distill key insights regarding the impact of annotations on fuzzing campaigns, including the finding that LLMs can substitute for human domain experts when generating annotations without significant performance degradation.
\end{itemize}

\section{Background}

We briefly introduce the necessary background.

\subsection{IJON}
\label{sec:background_ijon}

Fuzz testing has emerged as a dominant paradigm for automated software security analysis, typically relying on code coverage feedback to explore a program's state space.
However, coverage alone may fail to capture the subtle state transitions required to reach deeper program logic.

Consider a maze-solving program in Listing~\ref{lst:ijon_example}. A coverage-guided fuzzer can easily explore almost every branch, such as the movement cases (lines 6--9) and the boundary check (line 13), within a few iterations.
However, it will likely struggle to trigger the \texttt{Bug()} branch (line 11).
This difficulty arises because standard coverage provides no gradient. The fuzzer receives the same feedback signal whether the player is at the starting position or just one step away from the goal. Without a metric to differentiate progress, the fuzzer cannot distinguish a ``near-miss'' from a useless input.

\begin{listing}[h]
\begin{lstlisting}[style=aijonBase, language=C, basicstyle=\ttfamily\footnotesize]
while (true) {
    ox=x; oy=y;
    // Provide feedback about the current position
    IJON_SET(hash_int(x, y));
    switch (input[i]) {
        case 'w': y--; break;
        case 's': y++; break;
        case 'a': x--; break;
        case 'd': x++; break;
    }
    if (maze[y][x] == '#') { Bug(); }
    // If the target is blocked, do not advance
    if (maze[y][x] != ' ') { x = ox; y = oy; }
}
\end{lstlisting}
\caption{Example of IJON-style annotations for \textsc{maze}.}
\label{lst:ijon_example}
\end{listing}

To address this limitation, IJON proposes that developers manually insert annotations into the source code to expose internal variables, thereby providing richer feedback to the fuzzer.
In the maze example, an annotation that tracks the player's $(x, y)$ coordinates allows the fuzzer to recognize each new position as a distinct state.
By transforming the search space from a binary ``hit-or-miss'' into a navigable map, the fuzzer can identify and retain seeds that reach new areas of the maze. This effectively focuses the fuzzer's ``energy'' on seeds that demonstrate progress toward the target, rather than wasting cycles on inputs that fail to advance the internal state.

In addition to \texttt{IJON\_SET} used in the maze example, IJON provides a diverse suite of annotation primitives tailored to different program behaviors.
To track monotonic progress, IJON offers \texttt{IJON\_MAX}, \texttt{IJON\_MIN}, and \texttt{IJON\_INC}. For comparison and distance, developers can use \texttt{IJON\_CMP}, \texttt{IJON\_DIST}, and \texttt{IJON\_STRDIST}. By leveraging these manual hints, IJON successfully identified 10 vulnerabilities in the DARPA CGC dataset~\cite{cgc} that remained undiscovered by state-of-the-art tools like AFL, REDQUEEN~\cite{aschermann2019redqueen}, QSYM~\cite{yun2018qsym}, and T-Fuzz~\cite{peng2018t}.

\subsection{Large Language Models for Code Comprehension/Generation}


Recent advancements in large language models (LLMs) have significantly enhanced their ability to comprehend complex logic and generate syntactically correct code.
These improvements have led to the widespread adoption of LLMs in domains where deep semantic understanding is paramount, such as automated program repair. For instance, in DARPA's AI Cyber Challenge (AIxCC)~\cite{aixcc}, teams successfully leveraged LLMs to generate functional patches for both synthetic and zero-day vulnerabilities.

Specifically, the high-level reasoning exhibited by modern LLMs offers a viable path toward bridging the gap between manual instrumentation and automated testing. By automating the generation of IJON-style annotations, it becomes possible to reduce the traditional bottleneck of human domain expertise. This transition enables scalable annotation-based fuzzing, making large-scale evaluation and vulnerability detection feasible for complex software systems that were previously too labor-intensive to instrument.

\section{Preliminary Study}
\label{sec:motivation}

\begin{table}[t]
    \centering
    \small
    \caption{Median time to solve levels in \textit{Super Mario Bros.} and the solve-time ratio between LLM-generated annotations and the IJON authors' annotations. The geometric-mean ratio shows that LLM-generated annotations achieve performance comparable to the IJON authors' annotations.}
    \label{tab:ijon-aijon-median-ratio}
    \begin{tabular}{l c r c r r}
        \toprule
        Level & Solved & IJON & Solved & IJON+LLM & Ratio \\
        \midrule
        1-1  & \cmark & 42.23  & \cmark & 15.60     & 0.37 \\
        1-3  & \cmark & 32.95    & \cmark & 12.83  & 0.39 \\
        2-3  & \cmark & 53.70     & \cmark & 23.48   & 0.43 \\
        3-1 & \cmark & 83.08  & \cmark & 47.77  & 0.57 \\
        3-2 & \cmark & 47.55    & \cmark & 16.55    & 0.34 \\
        3-3 & \cmark & 11.95    & \cmark & 14.32  & 1.19 \\
        4-1 & \cmark & 13.10     & \cmark & 42.13  & 3.21 \\
        4-3 & \cmark & 16.73  & \cmark & 26.42  & 1.57 \\
        5-1 & \cmark & 24.75    & \cmark & 107.50    & 4.34 \\
        5-2 & \cmark & 25.42  & \cmark & 22.28  & 0.87 \\
        5-3 & \cmark & 10.58  & \cmark & 20.55    & 1.94 \\
        6-1 & \cmark & 18.45    & \cmark & 23.03  & 1.24 \\
        6-3 & \cmark & 18.93  & \cmark & 10.83  & 0.57 \\
        7-1 & \cmark & 28.50     & \cmark & 21.63  & 0.75 \\
        7-3 & \cmark & 17.57  & \cmark & 17.90     & 1.01 \\
        8-1 & \cmark & 289.28 & \cmark & 164.37 & 0.56 \\
        8-2 & 2/3 & 306.28  & 1/3 & 388.51      & 1.26 \\
        8-3 & \cmark & 27.10     & \cmark & 61.15    & 2.25 \\
        \midrule
        \multicolumn{5}{l}{\textbf{Geometric mean ratio}} & \textbf{0.96} \\
        \bottomrule
    \end{tabular}
\end{table}

In prior work, Aschermann et al.~\cite{aschermann2020ijon} manually inserted annotations into target programs to guide a fuzzer toward specific goals.
One of the target programs used in their evaluation was \textit{Super Mario Bros.}, and the fuzzer's objective was to complete levels in the game.
They demonstrated that inserting annotations into the source code of \textit{Super Mario Bros.} enabled the fuzzer to complete levels significantly faster than vanilla AFL.
This ability of annotations to guide the fuzzer toward specific goals demonstrates their effectiveness in improving fuzzer performance.
However, this approach does not scale well because it requires developer expertise to identify optimal locations and state-representing variables in the target program.

We observe that the target program's source code contains valuable information about its state, such as variables that can be leveraged to generate annotations.
Recent advancements in Large Language Models (LLMs) have significantly improved their ability to comprehend and generate syntactically correct code.

To quickly evaluate the potential of LLMs to generate annotations, we used an LLM to generate annotations for the IJON \textit{Super Mario Bros.} program.
We manually pasted the code into a prompt for the ChatGPT model \texttt{gpt-4.1} and asked it to generate annotations to guide a fuzzer to complete levels in the game.
The full prompt can be found in Listing \S~\ref{lst:supermariobros_prompt} in the Appendix.
We applied the LLM-generated annotations to the source code of the \textit{Super Mario Bros.} program and fuzzed it using the version of IJON published by the authors.
We conducted the experiment on an AWS EC2 instance with 380 Intel(R) Xeon(R) 6975P-C @ 2.7 GHz cores and 744 GB of RAM. Each trial ran for 8 hours, and we conducted 3 trials across 28 levels of the Mario game.
The fuzzers found crashes in 18 of the 28 levels.
We compare the median time to solve, across the three trials, for the 18 levels in Table~\ref{tab:ijon-aijon-median-ratio}.

We observe that the LLM-generated annotations provided performance comparable to the manually inserted annotations used in IJON.
This result demonstrates the potential of LLMs to automatically generate IJON-style annotations, eliminating the need for developer expertise.
\section{\aijon: Scaling IJON Annotation using LLMs}%
\label{sec:design}\label{sec:aijon}
Motivated by the results of our preliminary study, we aim to leverage the capabilities of LLMs to automatically generate IJON-style annotations for target programs.

To bridge the gap between the manual effort required for annotations and large-scale vulnerability detection, we present \aijon, a system that harnesses the natural language comprehension and code generation capabilities of large language models (LLMs) to automate the generation of IJON-style annotations.

\begin{figure*}[t]
    \centering
    \includegraphics[width=\textwidth]{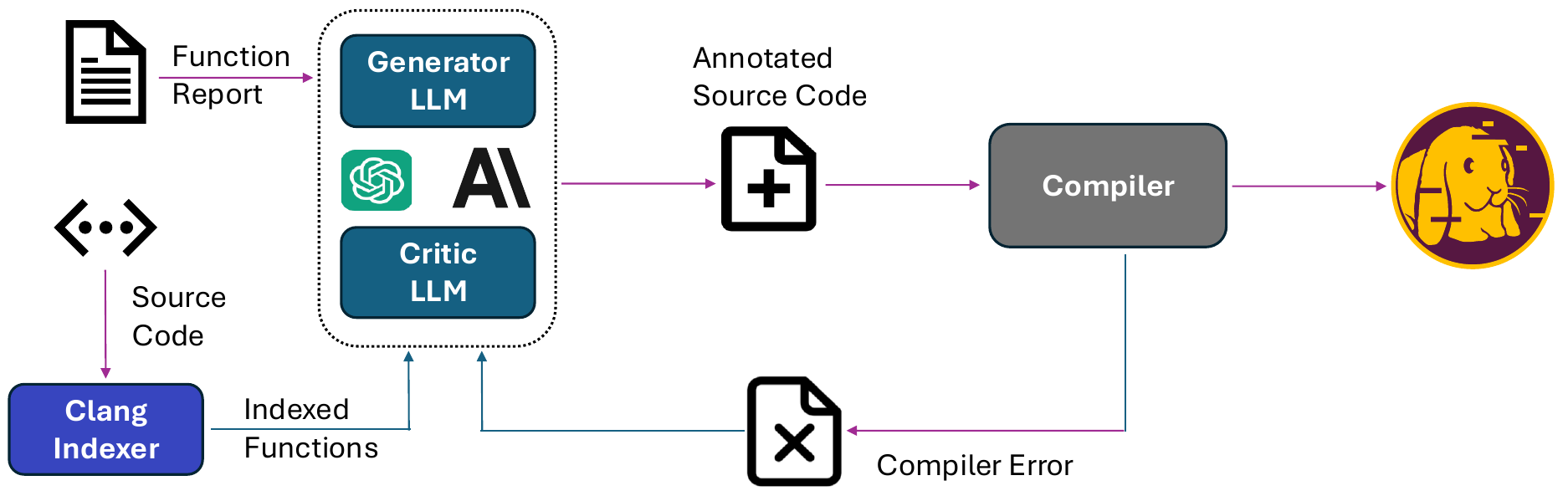}
    \caption{Design of \aijon.}
    \label{fig:design}
\end{figure*}

\begin{figure}[t]
    \centering
    \includegraphics[width=0.45\textwidth]{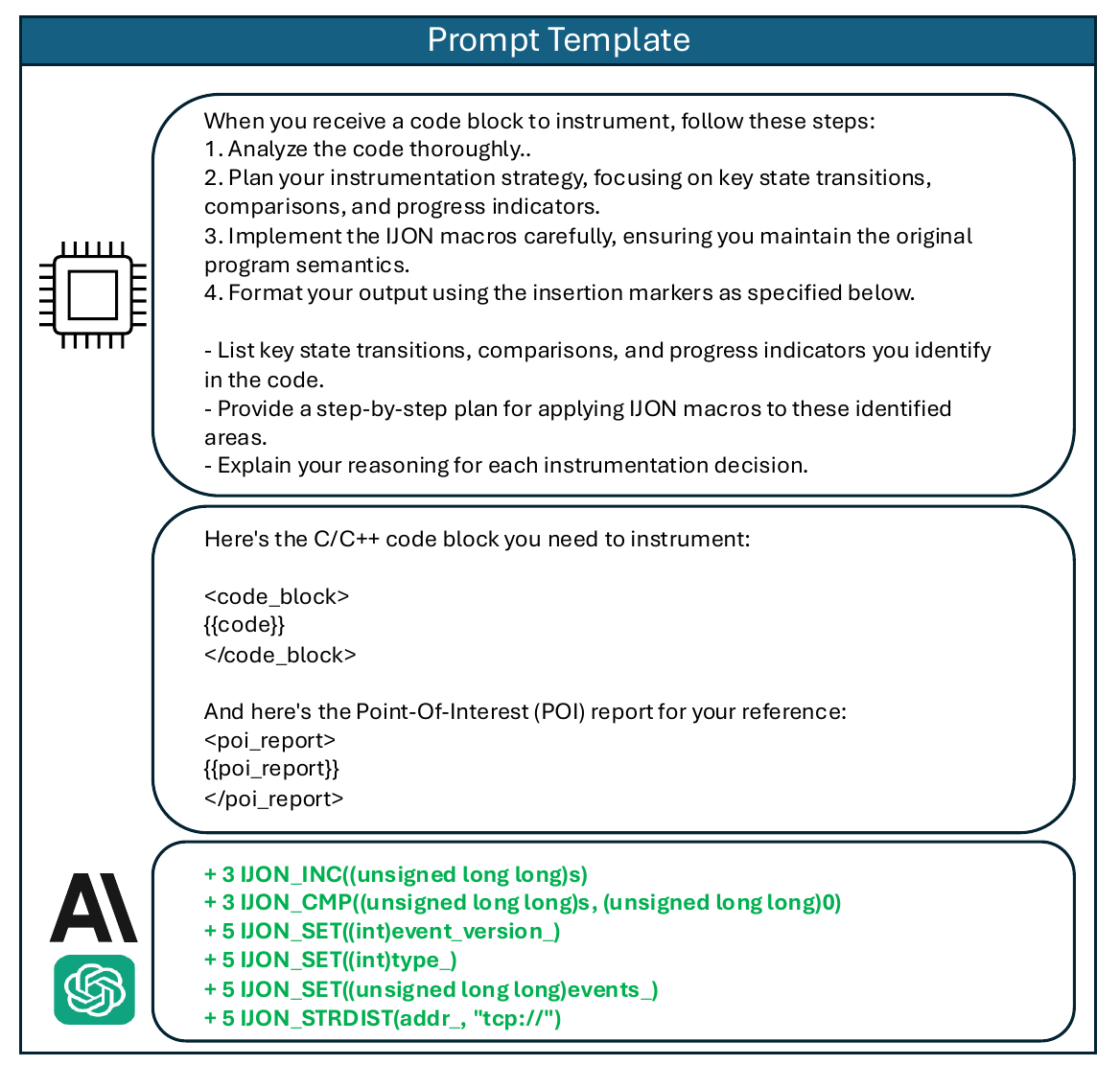}
    \caption{Example prompt for an LLM to generate IJON annotations.}
    \label{fig:prompt_example}
\end{figure}


\textbf{Overview.} Our primary goal in designing \aijon is to create a scalable and automated approach for generating IJON-style annotations for target programs.
An overview of \aijon is provided in Figure~\ref{fig:design}.
Given the source code and an externally generated (or user-supplied) list of potential annotation locations (a \emph{Function Report}), a generator LLM produces annotations, which a second LLM then critiques and refines.
The annotated code is then compiled with the fuzzer's instrumentation and is ready for fuzzing.
If compilation fails, we enter a bounded refinement loop and provide the generator LLM with the compiler error message.
Next, we describe each step in more detail.

\subsection{Design Considerations}
We first outline key design considerations.

\smallskip
\noindent
\textbf{Function Indexing.} Because LLMs have limited context windows, it is not feasible to feed the entire source code of a large project to the LLM to generate annotations.
Therefore, we rely on an indexer that is capable of parsing the source code of the target project and extracting individual functions from it.
Using this indexer, we can query individual functions in the target program and feed them to the LLM while staying within the context window.

\smallskip
\noindent
\textbf{Target Locations for Annotations.} Automatically identifying optimal locations for annotations is a non-trivial task and is outside the scope of this work.
As such, we rely on external analyses (e.g., static analysis tools) to provide a list of high-interest locations in the target program that should be annotated.
We call this list the \textit{Function Report}, as it is organized on a per-function basis.
We design \aijon to be flexible enough to handle different Function Report formats, such as YAML/JSON files produced by different static analyses or patch files that identify newly added lines in the target program.
We aim for a high degree of compatibility with a wide range of tools and formats, enabling ease of use.
Using the Clang indexer, \aijon can uniquely identify the correct function given the file path and line number of the annotation site.
Alternatively, \aijon can use a function index key (a unique identifier for each function generated by the indexer), if provided, to directly identify the function to be annotated.

\smallskip
\noindent
\textbf{LLM Planner-Critic Model.} To generate high-quality annotations, we adopted the planner-critic approach for annotation generation with LLMs~\cite{llm_planner_critic_model}.
This approach was found to produce higher quality annotations compared to a single-shot generation approach.
Since the planner agent is responsible for reading the full source code of the function to be annotated, we opted to use a model with a large context window, such as \texttt{gpt-4.1}.
For the critic agent, we used \texttt{gpt-o3}, which is designed for tasks that involve deep reasoning.

\smallskip
\noindent
\textbf{Prompt Design.} When designing the prompts for \aijon, we considered the key information that would be necessary for generating high-quality IJON-style annotations.
A human would typically analyze the source code of the function to be annotated and the static report to understand the goal of the annotation.
They would then identify key variables involved in reaching the goal and select appropriate IJON primitives to annotate these variables.

Similarly, we designed the prompts for \aijon to include the following guidelines:
\begin{itemize}[noitemsep, topsep=0pt]
    \item Consult a cheat sheet of IJON primitives and their descriptions.
    \item Analyze the provided source code of the function to be annotated to understand its logic.
    \item Use the provided Function Report to identify the parts of the function source code that are relevant to the annotation goal.
    \item Create a high-level plan for IJON annotations that would help a fuzzer reach the goal specified in the Function Report.
\end{itemize}

A shortened representation of our prompt design is shown in Figure~\ref{fig:prompt_example}. The full prompt design can be found in Listing~\ref{lst:magma_prompt} in the Appendix.

\smallskip
\noindent
\textbf{Runtime.} We re-implemented the techniques and primitives introduced by IJON on top of AFL++ v4.30c, enabling us to take advantage of modern fuzzing features and improvements.
We use this modified version of AFL++ (henceforth referred to as IJON) in our evaluations.
The authors of IJON recommend using a parallel configuration with two cores, where IJON runs alongside vanilla AFL++ and the two instances share interesting seeds~\cite{ijon_github}.
Following this recommendation, \aijon uses a similar parallel configuration for fuzzing the target program with the generated annotations.

With these considerations in mind, we implemented \aijon to automate the generation of IJON-style annotations using LLMs.
Our prompt integrates the crucial information about IJON primitives, the source code of the function to be annotated, and the Function Report. We now discuss \aijon's workflow in more detail.

\subsection{Workflow}

\aijon's workflow is illustrated in Figure~\ref{fig:design}.
Given a target project packaged as an OSS-Fuzz project~\cite{oss_fuzz_format} and a Function Report, \aijon begins its workflow.
\aijon first builds the target project and then uses Clang Indexer~\cite{clang_indexer} to parse the full source code of the target project and extract individual functions.
It then iterates over each entry in the Function Report.
For each entry, \aijon identifies the function to be annotated either by the combination of file path and line number or by using a function index key to look up the function in the indexer.
Once \aijon retrieves the source code of the function to be annotated, it queries the Generator LLM Agent with a prompt that includes the source code of the function and the Function Report.

The Generator LLM Agent analyzes the provided source code of the function and the Function Report entry to understand the task.
Using the IJON cheat sheet as a reference for IJON primitives, it generates a list of IJON-style annotations that would help a fuzzer reach the goal specified in the Function Report.

This list of annotations is then passed to the Critic LLM Agent, which reviews the annotations and attempts to improve them by removing redundant or syntactically incorrect annotations.
After this review, \aijon performs several static checks on the annotations to ensure that they do not introduce any syntax errors or unintended side effects in the target program.
For example, any annotation that invokes a function is removed during this filtering step to ensure that no side effects are introduced in the target program due to the annotations.

\begin{listing}[h]
\begin{lstlisting}[style=aijonBase, language=C]
IJON_SET((int)((mask & info_ptr->free_me & PNG_FREE_EXIF) == 0)); /* PATCHID:15256 */
IJON_CMP((unsigned long long)(mask & info_ptr->free_me), (unsigned long long)PNG_FREE_EXIF); /* PATCHID:14556 */
IJON_SET((int)(info_ptr->eXIf_buf != NULL)); /* PATCHID:14156 */
    MAGMA_LOG("PNG006", MAGMA_AND(info_ptr->eXIf_buf != NULL, (mask & info_ptr->free_me & PNG_FREE_EXIF) == 0));
\end{lstlisting}
\caption{An example of annotations generated by \aijon for PNG006 in the Magma dataset.}%
\label{lst:aijon_example_patch}
\end{listing}

Once the annotations have been reviewed and passed the static checks, \aijon applies the annotations to the source code of the target program.
It then attempts to compile the target program with the new source code.

If any errors are encountered during compilation, \aijon invokes the LLM again to review the annotations and resolve the compilation errors.
This process is repeated until the target program compiles successfully with the new annotations or a maximum number of attempts is reached.\footnote{Currently we set this maximum number of attempts to 10.}

Once the target program has been successfully compiled with the new annotations, \aijon starts fuzzing the target program's harnesses.
We use a two-instance setup, with one instance running IJON and the other running vanilla AFL++; interesting test cases are shared using AFL++'s native synchronization mechanism.

\section{Evaluation of Annotations}%
\label{sec:aijon_magma}


To evaluate the impact of automated IJON annotations on real-world fuzzing, we benchmarked \aijon against AFL++ using the Magma dataset.
While previous sections detailed the LLM-driven workflow for generating these annotations, this evaluation focuses on \aijon's performance at scale.
Specifically, we aim to determine if automated annotations can effectively reduce vulnerability survival time and enhance detection capabilities in complex, real-world software.

\smallskip
\noindent
\textbf{Dataset.} The Magma benchmark~\cite{Hazimeh:2020:Magma} comprises 138 vulnerabilities across nine real-world C/C++ projects.
This dataset contains the source code of these projects that have been modified to include injected vulnerabilities which are similar to real-world vulnerabilities that were discovered in these projects in the past.
These vulnerabilities are introduced via source-code patches injected during compilation.
The patches that implement specific bugs are explicitly labeled with unique vulnerability IDs.

In each project, the vulnerabilities are injected with a \texttt{MAGMA\_LOG} statement.
This instrumentation allows precise tracking of when the vulnerable code is first reached during fuzzing.
The arguments to this \texttt{MAGMA\_LOG} statement are conditional statements that use variable values which is used to determine if the conditions for the vulnerability are met.
When the fuzzer generates an input that satisfies these conditions, the vulnerability is triggered and Magma is also able to precisely track the first time the vulnerability is triggered.
This instrumentation allows evaluating fuzzers based on their ability to reach and trigger these vulnerabilities.

This instrumentation can be leveraged to determine locations in the target program for inserting IJON-style annotations.
Since the vulnerability is detected only when the \texttt{MAGMA\_LOG} statement is reached, a good location for the annotations would be around this statement.
Furthermore, since the conditions necessary for triggering the vulnerability can be obtained from the arguments to this \texttt{MAGMA\_LOG} statement, this information can be leveraged to generate the corresponding annotations that help the fuzzer trigger these vulnerabilities faster.




We augmented the Magma suite to include AFL++ v4.30c as a supported fuzzer, so as to ensure a fair comparison between \aijon and AFL++.
At the conclusion of each trial, Magma generates detailed reports identifying reached and triggered vulnerabilities, alongside their respective time-to-reach and time-to-trigger metrics.
For the purposes of this evaluation, a fuzzer is credited with \textit{reaching} a vulnerability if it does so in at least five of its ten independent 24-hour runs.

\begin{table}[h]
    \centering
    \caption{Distribution of vulnerabilities in Magma dataset and those reached by AFL++ within 24 hours.}%
    \label{tab:magma_vuln_distribution}
    {\normalsize
    \begin{tabular}{ccc}
        \toprule
        \textbf{Project} & \textbf{Vulnerabilities} & \textbf{Reached by AFL++} \\
        \midrule
        Lua & 4 & 3 \\
        PHP & 16 & 5 \\
        Poppler & 22 & 15 \\
        LibPNG & 7 & 6 \\
        SQLite3 & 20 & 15 \\
        OpenSSL & 20 & 10 \\
        LibTIFF & 14 & 10 \\
        LibXML2 & 17 & 8 \\
        Libsndfile & 18 & 0 \\
        \bottomrule
        Total & 138 & \reachedmagmabugs \\
        \bottomrule
    \end{tabular}
    }
\end{table}

To establish our evaluation subset, we ran AFL++ on the full Magma suite for 24 hours. At the conclusion of this period, AFL++ successfully reached \reachedmagmabugs vulnerabilities\footnote{For the project, Libsndfile, the sndfile\_fuzzer did not find any new inputs. Therefore, AFL++ was not able to reach any vulnerabilities in this project within the time frame.}, as shown in Table~\ref{tab:magma_vuln_distribution}.
Following our methodology of isolating the bug triggering performance, we use these \reachedmagmabugs reached vulnerabilities covering eight projects as the primary targets for our comparative evaluation.

\smallskip
\noindent
\textbf{Baselines.} We evaluate two configurations to evaluate annotations on a large scale:

\begin{itemize}[nosep, leftmargin=*]
    \item \textbf{AFL++}: AFL++ v4.30c in a standard parallel configuration consisting of one main and one secondary instance that use the native AFL++ synchronization mechanism to share interesting test cases.
    \item \textbf{\aijon}: This configuration pairs one main instance utilizing \aijon-generated annotations with a secondary AFL++ v4.30c instance. Both instances share interesting test cases using the native AFL++ synchronization mechanism.
    To facilitate automatic annotation, we provide \aijon with a single Function Report for each project consisting of all relevant vulnerability-injection patches.
    \aijon then automatically annotates all functions modified by these patches. We manually verified that \aijon successfully generated valid IJON-style annotations for \aijonmagmaannotated out of \reachedmagmabugs vulnerabilities, representing a 98.6\% success rate. In the single failing case, \aijon was unable to produce valid annotations within the maximum attempt limit.
\end{itemize}

\smallskip
\noindent
\textbf{Experiment Setup.}
Throughout our evaluation, each configuration is allocated two cores for fuzzing a single target program.
Each experiment runs ten times for a total of 24 hours, with results averaged over the 10 independent trials to ensure statistical significance and reduce the impact of the randomness.
All experiments were conducted on a Kubernetes cluster of Intel(R) Xeon(R) CPU E5-2670 v2 @ 2.50GHz.
Each Kubernetes Pod was allocated 20 cores and 2 GB of RAM for running a set of 10 trials of a fuzzing compaign for a single target program and was running Ubuntu 18.04 with Linux 6.8.0 64-bit.
We used the seeds that were provided by the Magma benchmark for each program as the initial seed corpus for our experiments.

\smallskip
\noindent
\textbf{Metrics.}
We focus our evaluation on two primary metrics:
\begin{itemize}[nosep, leftmargin=*]
    \item \textit{Vulnerabilities Triggered}: The total number of unique vulnerabilities successfully triggered within the 24-hour fuzzing period.
    \item \textit{Survival Time}: The time taken to trigger a vulnerability, measured from the first timestamp when this vulnerability was reached until the vulnerability is triggered.
\end{itemize}
Since \aijon places annotations only within the functions containing vulnerabilities, we do not evaluate the time taken to reach a vulnerability, as the annotations do not impact this metric.
Furthermore, once a vulnerable location is reached, there is typically little code-coverage that can guide a fuzzer towards triggering the vulnerability.
Therefore, the annotations can play a significant role in guiding the fuzzer towards triggering the vulnerability once it has been reached.
With this understanding, comparing the survival times of vulnerabilities provides a clear evaluation of the impact of IJON-style annotations in guiding fuzzers towards triggering vulnerabilities.

\subsection{Large-Scale Experiment on Magma}


\paragraph{Experiment.} We now attempt to replicate IJON's findings across a larger dataset in the context of vulnerability-finding, namely the Magma benchmark.
To this end, we use our aforementioned experimental setup and conduct a large-scale analysis of the impact of annotations by running \aijon and AFL++ ten times for 24-hours on the nine projects of the Magma benchmark.
For \aijon, each target has received annotations a priori.

\paragraph{Hypothesis.} Our hypothesis is that IJON-style annotations enhance fuzzing by providing targeted feedback which can be observed through reduced survival times and an increased number of vulnerabilities triggered.
In this experiment, we expect to observe that \aijon would be able to trigger more vulnerabilities than AFL++ within the 24-hour fuzzing period and that it would be able to trigger vulnerabilities faster than AFL++ due to the additional guidance provided by the IJON-style annotations.
However, our large-scale analysis shows that the impact of annotations is more nuanced than commonly assumed in the fuzzing community, with our experimental results containing both positive and negative findings.

\begin{figure}[t]
    \centering
    \includegraphics[width=0.5\textwidth]{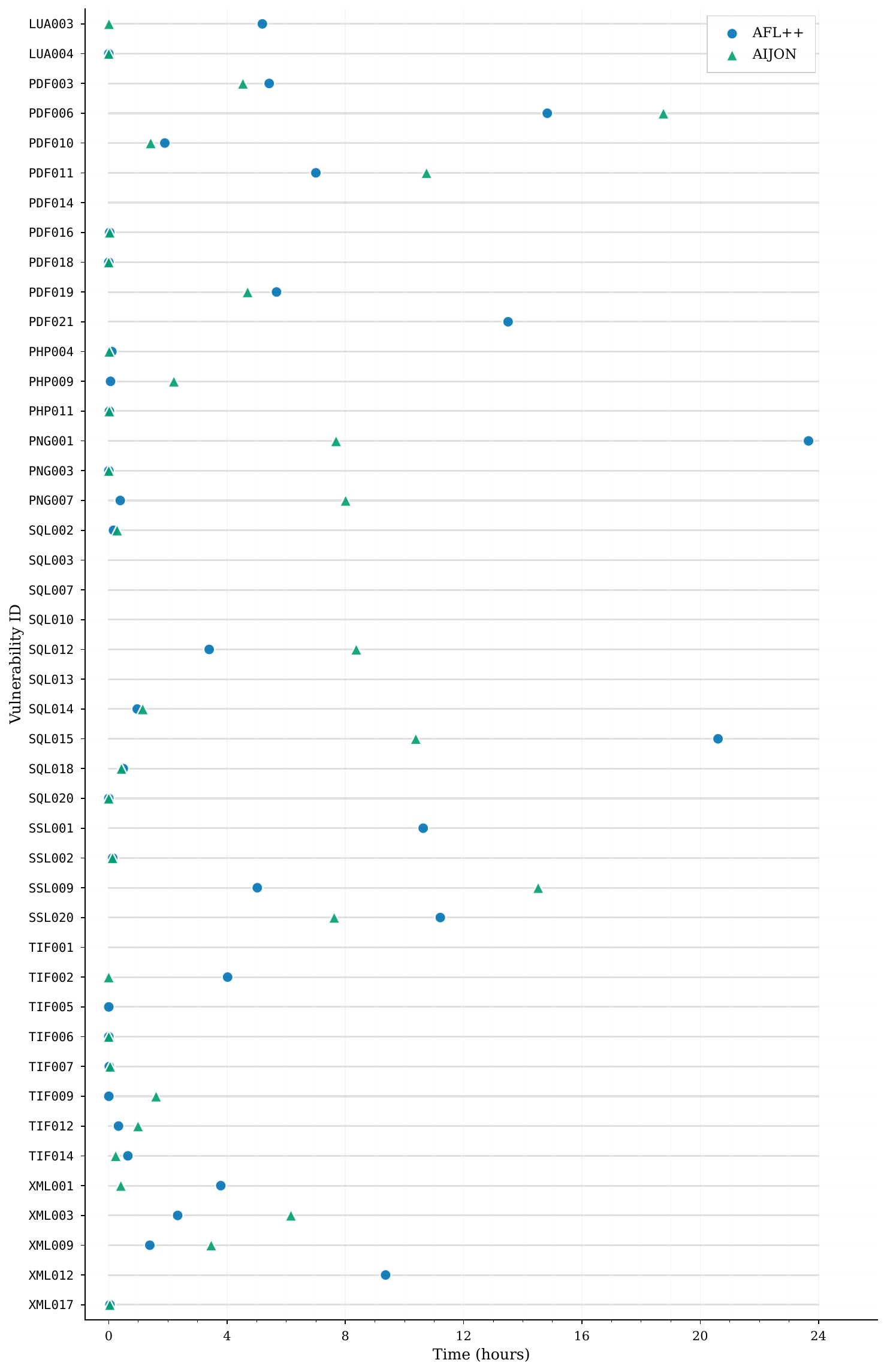}
    \caption{Survival times of vulnerabilities triggered by AFL++ and \aijon on Magma dataset.}
    \label{fig:magma_survival_times}
\end{figure}

\paragraph{Result analysis \& findings.} After 24 hours of fuzzing, AFL++ was able to trigger \aflppmagmatriggered vulnerabilities across 10 trials.
In the same time, \aijon was able to trigger \aijonmagmatriggered vulnerabilities, missing \aijonmissed vulnerabilities that were triggered by AFL++.
Among the vulnerabilities that were triggered by \aijon and AFL++, we observed that \aijon was able to trigger \aijonfaster vulnerabilities faster when compared to AFL++ with an average speedup of 37.82\%.
The survival times for each of the vulnerabilities are plotted in Figure~\ref{fig:magma_survival_times}.


\begin{table}[tb]
    \centering
    \footnotesize
	\caption{Comparison of vulnerabilities triggered by AFL++, and \aijon in Magma dataset}
    \label{table:magma_survival_times}
    \begin{tabular}{cccccc}
        \toprule
		\textbf{BUG ID} & \textbf{AFL++} & \textbf{\aijon} & \textbf{BUG ID} & \textbf{AFL++} & \textbf{\aijon} \\
        \midrule
        LUA003 & \cmark & \cmark & SQL015 & \cmark & \cmark \\
        LUA004 & \cmark & \cmark & SQL018 & \cmark & \cmark \\
        PDF003 & \cmark & \cmark & SQL020 & \cmark & \cmark \\
        PDF006 & \cmark & \cmark & SSL001 & \cmark & \xmark \\
        PDF010 & \cmark & \cmark & SSL002 & \cmark & \cmark \\
        PDF011 & \cmark & \cmark & SSL009 & \cmark & \cmark \\
        PDF014 & \xmark & \xmark & SSL020 & \cmark & \cmark \\
        PDF016 & \cmark & \cmark & TIF001 & \xmark & \xmark \\
        PDF018 & \cmark & \cmark & TIF002 & \cmark & \cmark \\
        PDF019 & \cmark & \cmark & TIF005 & \cmark & \xmark \\
        PDF021 & \cmark & \xmark & TIF006 & \cmark & \cmark \\
        PHP004 & \cmark & \cmark & TIF007 & \cmark & \cmark \\
        PHP009 & \cmark & \cmark & TIF009 & \cmark & \cmark \\
        PHP011 & \cmark & \cmark & TIF012 & \cmark & \cmark \\
        PNG001 & \cmark & \cmark & TIF014 & \cmark & \cmark \\
        PNG003 & \cmark & \cmark & XML001 & \cmark & \cmark \\
        PNG007 & \cmark & \cmark & XML003 & \cmark & \cmark \\
        SQL002 & \cmark & \cmark & XML009 & \cmark & \cmark \\
        SQL012 & \cmark & \cmark & XML012 & \cmark & \xmark \\
        SQL013 & \xmark & \xmark & XML017 & \cmark & \cmark \\
        SQL014 & \cmark & \cmark &  &  & \\
        \bottomrule
    \end{tabular}
\end{table}

Upon first glance at the results, we found that \aijon's results on the Magma benchmark were mixed.
In \aijonfaster cases, \aijon had a clear advantage of AFL++ in terms of survival times, while in other cases, AFL++ outperformed \aijon.
Furthermore, \aijon had failed to trigger \aijonmissed vulnerabilities that were triggered by AFL++ which is a significant drawback.

\medskip
In summary, the annotations did not provide a significant improvement over the baseline AFL++ as expected. In fact, they led to less bugs found, however, provided a speedup in finding some bugs.

\subsection{Dissecting the Performance Results}%
\label{sec:ijon_magma_results}


To understand how annotations were affecting these results and exclude experimental error, we performed further experiments to isolate the cause of these negative performances of \aijon.
We develop two possible hypotheses that could explain these results:
\begin{itemize}[nosep, leftmargin=*]
    \item \textbf{H1: Low Quality Annotations}: Since \aijon relies on LLMs to generate IJON-style annotations, there is a possibility that the LLM may not always generate high quality annotations that a human expert would have generated. This could affect the overall performance of \aijon on the Magma dataset.
    \item \textbf{H2: Annotation Interference}: Due to the presence of multiple vulnerabilities in the same target program, the annotations could interfere with each other and lead to sub-optimal performance.
\end{itemize}

To test these hypotheses, we designed two experiments to isolate the impact of each of these factors on \aijon's performance on the Magma dataset.

\begin{figure}[t]
    \centering
    \includegraphics[width=0.45\textwidth]{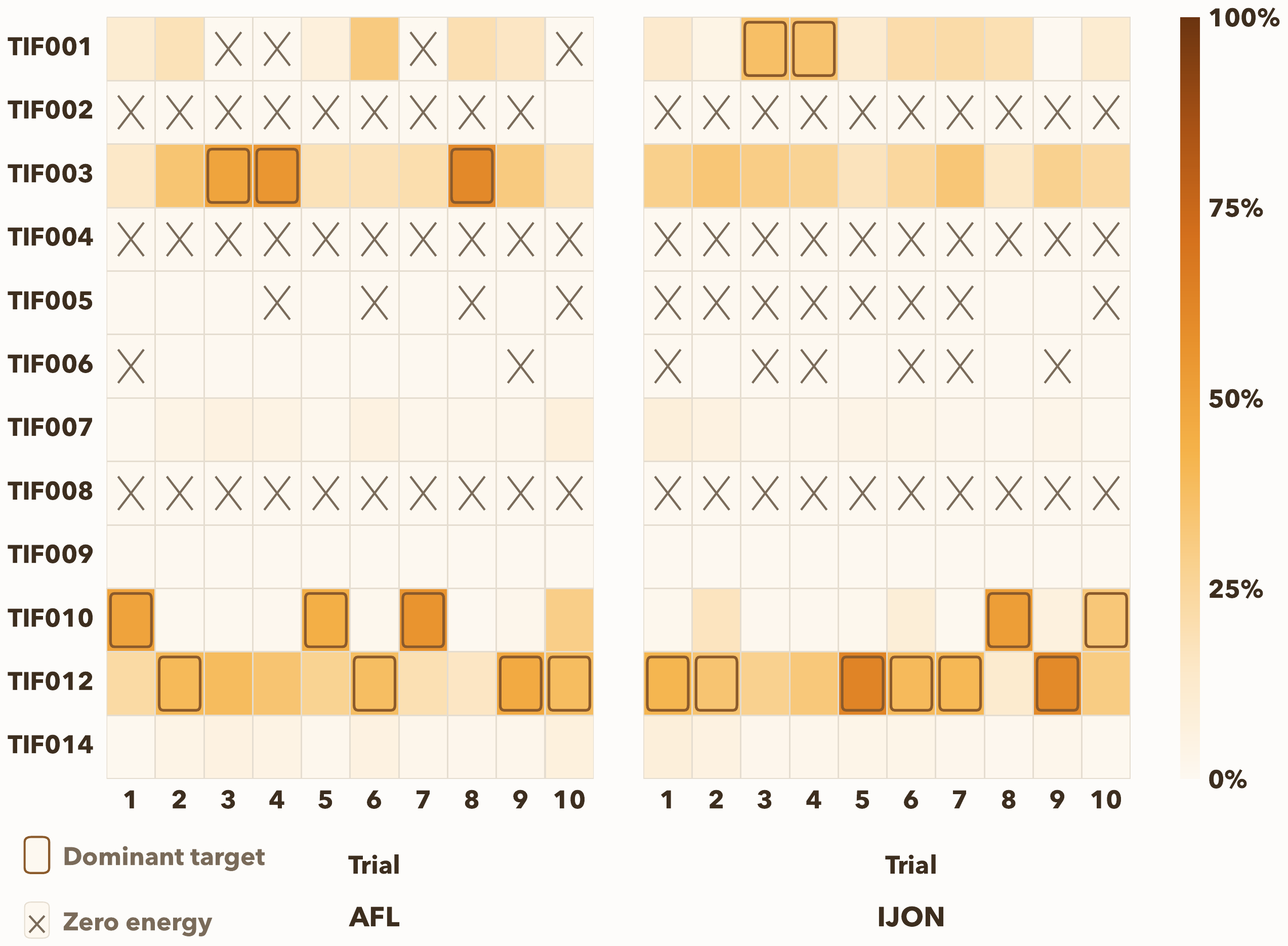}
    \caption{AFL++ vs IJON runtime fuzzing energy distribution per trial for \texttt{tiffcp}.}
    \label{fig:tiffcp_heatmap}
\end{figure}

\subsection{H1: Manually Generated Annotations}
\begin{table}[tb]
    \centering
    \footnotesize
    \caption{Comparison of vulnerabilities triggered by AFL++, IJON and \aijon in Magma dataset}
    \label{table:magma_vuln_comparison}
    {\setlength{\tabcolsep}{3pt}
    \begin{tabular}{cccccccc}
        \toprule
		\textbf{BUG ID} & \textbf{AFL++} & \textbf{IJON} & \textbf{\aijon} & \textbf{BUG ID} & \textbf{AFL++} & \textbf{IJON} & \textbf{\aijon} \\
        \midrule
        LUA003 & \cmark & \cmark & \cmark & SQL015 & \cmark & \cmark & \cmark \\
        LUA004 & \cmark & \cmark & \cmark & SQL018 & \cmark & \cmark & \cmark \\
        PDF003 & \cmark & \cmark & \cmark & SQL020 & \cmark & \cmark & \cmark \\
        PDF006 & \cmark & \cmark & \cmark & SSL001 & \cmark & \xmark & \xmark \\
        PDF010 & \cmark & \cmark & \cmark & SSL002 & \cmark & \cmark & \cmark \\
        PDF011 & \cmark & \cmark & \cmark & SSL009 & \cmark & \cmark & \cmark \\
        PDF014 & \xmark & \cmark & \xmark & SSL020 & \cmark & \cmark & \cmark \\
        PDF016 & \cmark & \cmark & \cmark & TIF001 & \xmark & \xmark & \xmark \\
        PDF018 & \cmark & \cmark & \cmark & TIF002 & \cmark & \cmark & \cmark \\
        PDF019 & \cmark & \cmark & \cmark & TIF005 & \cmark & \cmark & \xmark \\
        PDF021 & \cmark & \cmark & \xmark & TIF006 & \cmark & \cmark & \cmark \\
        PHP004 & \cmark & \cmark & \cmark & TIF007 & \cmark & \cmark & \cmark \\
        PHP009 & \cmark & \cmark & \cmark & TIF009 & \cmark & \cmark & \cmark \\
        PHP011 & \cmark & \cmark & \cmark & TIF012 & \cmark & \cmark & \cmark \\
        PNG001 & \cmark & \cmark & \cmark & TIF014 & \cmark & \cmark & \cmark \\
        PNG003 & \cmark & \cmark & \cmark & XML001 & \cmark & \cmark & \cmark \\
        PNG007 & \cmark & \cmark & \cmark & XML003 & \cmark & \xmark & \cmark \\
        SQL002 & \cmark & \cmark & \cmark & XML009 & \cmark & \cmark & \cmark \\
        SQL012 & \cmark & \cmark & \cmark & XML012 & \cmark & \xmark & \xmark \\
        SQL013 & \xmark & \cmark & \xmark & XML017 & \cmark & \cmark & \cmark \\
        SQL014 & \cmark & \cmark & \cmark & & & \\
        \bottomrule
    \end{tabular}}
\end{table}

\begin{figure}[t]
    \centering
    \includegraphics[width=0.5\textwidth]{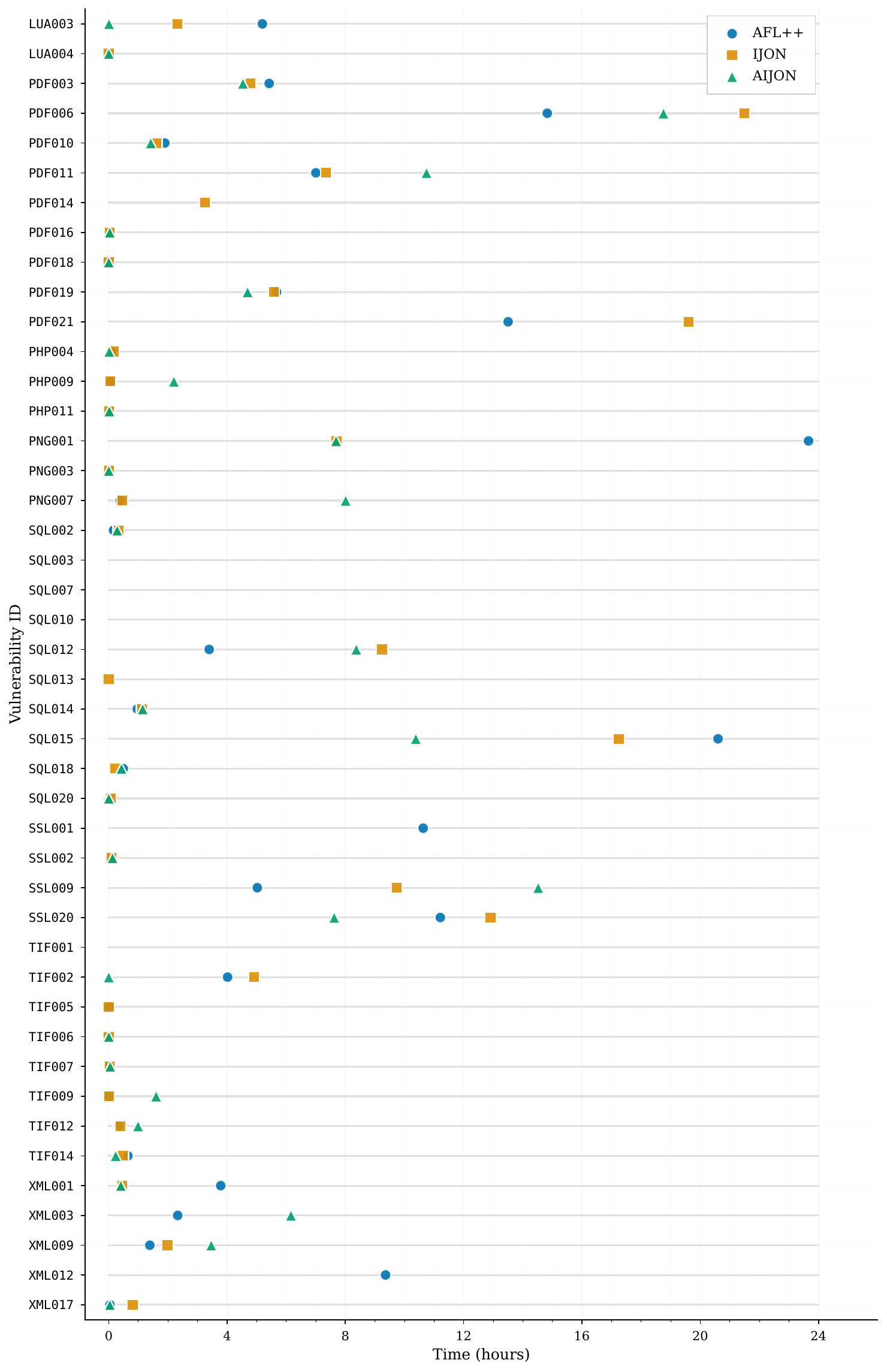}
    \caption{Survival times of vulnerabilities triggered by AFL++, \aijon and IJON on Magma dataset.}
    \label{fig:magma_ijon_survival_times}
\end{figure}

In our preliminary study ~\S\ref{sec:motivation}, we observed that LLM-generated annotations provided comparable performance to human-generated annotations on the \textit{Super Mario Bros.} program from IJON.
However, it is possible that real-world applications such as those present in the Magma dataset may be too complex for LLMs to generate high-quality annotations.

\paragraph{Experiment.} In order to test this hypothesis, we manually created IJON-style annotations for the \reachedmagmabugs vulnerabilities that were reached by AFL++ in our experiments from Section \S\ref{sec:aijon_magma}.
If the manually generated annotations also exhibit similar mixed results, this would indicate that IJON annotations inherently have a nuanced impact on fuzzing performance.
We then repeated the experiments from Section \S\ref{sec:aijon_magma} using these manually created annotations.

\paragraph{Result analysis \& findings}
Our hypothesis is that if the mixed performance results were due to low-quality LLM-generated annotations, then using manually generated annotations should lead to a significant improvement in performance.
According to the results, out of the \reachedmagmabugs vulnerabilities in our evaluation dataset, IJON successfully triggered \ijonmagmatriggered vulnerabilities.
However, IJON missed three vulnerabilities that were detected by AFL++.
At the same time, IJON successfully triggered two vulnerabilities that remained undiscovered by AFL++.
A detailed comparison of the vulnerabilities triggered by each configuration is provided in Table~\ref{table:magma_vuln_comparison}.

To evaluate efficiency, we further compared the survival times for the \ijonaflppcommon vulnerabilities triggered by both configurations (Figure~\ref{fig:magma_ijon_survival_times}).
Our analysis reveals that IJON achieved a faster time-to-trigger for \ijonfasteraflpp of these vulnerabilities compared to AFL++ with an average speedup of 34.34\%.

These results indicate that even with manually generated annotations, the survival times exhibit a mixed pattern similar to that observed with \aijon.
This suggests that LLM generated annotations are not the primary cause of the mixed performance results observed in Section \S\ref{sec:aijon_magma}.

\begin{tcolorbox}[colback=gray!10,colframe=black!75,title=Key Observation]
    LLMs can be used to generate IJON-style annotations thus reducing the need for manual effort allowing for automated and scalable generation of annotations without a significant degradation in performance.
\end{tcolorbox}

\subsubsection{Analysis of IJON on Magma}
\label{sec:ijon_magma_analysis}
In order to understand the multi-faceted effect that annotations were having on the fuzzer's ability to trigger vulnerabilities, we investigate the fuzzing process for all the vulnerabilities that were triggered by either AFL++ or IJON in detail.

\smallskip
\noindent
\textbf{Analysis metrics.} We utilize the fine-grained metrics that Magma provides such as the first time stamps when a vulnerability was reached and triggered as well as the number of times a vulnerability was reached or triggered during the full 24 hours of fuzzing.
These metrics allow fine-grained introspection into the fuzzing process and help us understand how annotations affect the fuzzer's behavior.

As we discussed in Section \S\ref{sec:background_ijon}, annotations allow fuzzers to distinguish between inputs that have similar edge coverage, but differ in terms of the values of certain variables that are annotated.
Using this additional feedback, the fuzzer is then motivated to explore more inputs that generate different values for these variables.
In essence, using annotations helps control the energy that the fuzzer distributes to different inputs and thereby different parts of the codebase.
And by focusing more energy on specific parts of the codebase, the fuzzer has higher chances of triggering vulnerabilities that are located in these parts of the codebase.

With this understanding, it would stand to reason that adding annotations to locations with vulnerabilities would always lead to a reduced survival time for these vulnerabilities.
A fuzzer would distribute more of its energy towards the locations that contain annotations and thus have more chances of triggering the vulnerabilities located there.
In theory, this should mean that IJON should always outperform AFL++ in terms of survival times for vulnerabilities that have annotations inserted.
However, our results indicate that the effect of annotations is more nuanced than this simplistic understanding.

The energy distributed by the fuzzer is a limited resource when fuzzing for 24 hours.
Since we annotated multiple vulnerabilities in the same codebase, this lead to contention for the energy of the fuzzer.
When one portion of the codebase receives significantly more energy than other parts, due to the presence of annotations, it changes the overall energy distribution across the codebase.
Since the fuzzer also saves interesting inputs to the queue, this change in energy distribution can also affect the areas that are targeted by the fuzzer in the future since the saved inputs are biased towards specific parts of the codebase.
Furthermore, this change in energy distribution can, at worst, lead to some parts of the codebase being ignored by the fuzzer altogether.

We observe this phenomenon when comparing the results of AFL++ and IJON on Magma.

XML003 was missed by IJON even though it was detected by AFL++.
We compared the number of times XML003 was reached by AFL++ and IJON after 24 hours of fuzzing and calculated the average value across 10 trials.
We found that AFL++ had reached this vulnerability almost 7 million times whereas IJON had only reached it almost 400,000 times.
Due to this difference in the reaching count, which represents the energy allocated by the fuzzer, IJON had far fewer chances of triggering this vulnerability compared to AFL++.
Instead, we found that IJON had allocated more energy towards XML012.

When a fuzzer distributes more energy towards specific parts of the codebase, it may continue to allocate energy even after the vulnerability has been triggered.
We found that TIF012 was triggered for the first time by IJON after it was reached 4 million times.
However, at the end of the 24 hours of fuzzing, IJON had reached TIF012 almost 64 million times.
This indicates that annotations continue to contend for the fuzzer's energy even after the vulnerability has been triggered.

A heatmap showing the energy distribution across different vulnerabilities in the LibTIFF project with the \texttt{tiffcp} binary is shown in Figure~\ref{fig:tiffcp_heatmap}.

\begin{tcolorbox}[colback=gray!10,colframe=black!75,title=Key Takeaway]
    In a multi-goal scenario, annotations can lead to contention for fuzzing energy, potentially reducing the energy allocated to other parts of the codebase and thereby affecting the fuzzer's ability to trigger other vulnerabilities.
\end{tcolorbox}

Beyond affecting the number of vulnerabilities that are triggered, annotations can also affect the time taken to trigger vulnerabilities.
One would assume that adding annotations to vulnerabilities that are reached very quickly would lead to a reduced survival time for these vulnerabilities.
However, we found that out of the 34 vulnerabilities that were triggered by both AFL++ and IJON, AFL++ was able to trigger 19 vulnerabilities faster than IJON.
Of those, AFL++ was finding 9 vulnerabilities faster than IJON by more than 1 hour.

Assume a situation where one target application has two vulnerabilities, A and B.
Vulnerability A is reached from one of the initial seed inputs S1.
Vulnerability B is reached from another initial seed input S2.
When the process of fuzzing starts, the fuzzer iterates over the initial seed corpus and therefore reaches both vulnerabilities A and B very quickly.
At this point, if the annotations inserted at vulnerability A win the competition for the fuzzer's energy, then the fuzzer will spend more time exploring inputs that reach vulnerability A.
As we have seen, this process can continue even after vulnerability A has been triggered.
Eventually, when the fuzzer has exhausted its energy for vulnerability A, it may then start exploring inputs that reach vulnerability B.
When the fuzzer eventually triggers vulnerability B, the survival time for vulnerability B would be longer than that of vulnerability A despite both vulnerabilities being reached very quickly.

In order to verify if this situation happens during our evaluation of IJON on Magma, we identified the first time stamp recorded by Magma when a vulnerability was triggered during fuzzing.
Using this information, we collected the number of times this same vulnerablity was reached during fuzzing.
Thus, by using the number of times the vulnerability was reached before it was triggered for the first time, we get a better idea of the impact of annotations on the fuzzer's ability to trigger vulnerabilities.

In the case for SSL009 where the survival time difference is significantly in favor of AFL++ (5 and 9 hours respectively), we found that IJON was able to trigger these vulnerabilities with a smaller value for the number of times it was reached.
This indicates that when the fuzzer allocated enough energy towards these vulnerabilities, it was able to trigger them quickly.
Furthermore, we found that IJON was instead allocating significantly more energy to SSL016.
After 24 hours of fuzzing and averaged across 10 trials, AFL++ had reached SSL016 almost 18 million times compared to IJON which had reached the same vulnerability over 1 billion times.
This subsequently led to reduced energy being allocated towards SSL009 (519,547 times by AFL++ vs 272,911 times by IJON) and thus extended the survival time for this vulnerability.

\begin{tcolorbox}[colback=gray!10,colframe=black!75,title=Key Takeaway]
    The change in energy distribution caused by annotation can delay detection of vulnerabilities even if they are easy to reach.
\end{tcolorbox}

\subsection{Single Goal Annotation}
\begin{table}[tb]
    \centering
    \footnotesize
	\caption{Comparison of vulnerabilities triggered by IJON and \aijon in Single-Goal Annotation compared to IJON on Multi-Goal Annotation in Magma dataset.}
    \label{table:single_magma_vuln_comparison}
    \setlength{\tabcolsep}{3pt}
    \begin{tabular}{ccccc}
        \toprule
		\textbf{BUG ID} & \textbf{\aijon} & \textbf{\aijon-SG} & \textbf{IJON} & \textbf{IJON-SG} \\
        \midrule
        SQL002 & \cmark & \cmark & \cmark & \cmark \\
        SQL003 & \xmark & \cmark & \xmark & \cmark \\
        SQL012 & \cmark & \cmark & \cmark & \cmark \\
        SQL013 & \xmark & \cmark & \cmark & \xmark \\
        SQL014 & \cmark & \cmark & \cmark & \cmark \\
        SQL015 & \cmark & \cmark & \cmark & \cmark \\
        SQL018 & \cmark & \cmark & \cmark & \cmark \\
        SQL020 & \cmark & \cmark & \cmark & \cmark \\
        \bottomrule
    \end{tabular}
\end{table}

\begin{table*}[tb]
    \centering
    \footnotesize
	\caption{Comparison of survival times and success rate for vulnerabilities in IJON-SG compared to IJON and \aijon-SG compared to \aijon in Magma dataset.}
    \label{table:single_magma_survival_comparison}
    \setlength{\tabcolsep}{3pt}
    \begin{tabular}{c|cc|cc|cc|cc}
        \toprule
		\textbf{BUG ID} & \multicolumn{2}{c|}{\textbf{\aijon}} & \multicolumn{2}{c|}{\textbf{\aijon-SG}} & \multicolumn{2}{c|}{\textbf{IJON}} & \multicolumn{2}{c}{\textbf{IJON-SG}} \\
        \midrule
        & \textbf{Time(s)} & \textbf{Success(\%)} & \textbf{Time(s)} & \textbf{Success(\%)} & \textbf{Time(s)} & \textbf{Success(\%)} & \textbf{Time(s)} & \textbf{Success(\%)} \\
        \midrule
        SQL002 & 1023 & 100 & 1276.5 & 100 & 1185 & 100 & 1239 & 100 \\
        SQL003 & - & 0 & 13135 & 33.33 & - & 0 & 53190 & 16.67 \\
        SQL012 & 30131.67 & 37.5 & 49685 & 20 & 33271.67 & 60 & 35056 & 50 \\
        SQL013 & - & 0 & 0 & 25 & 0 & 28.57 & - & 0 \\
        SQL014 & 4132.5 & 100 & 5299.5 & 100 & 4048.5 & 100 & 4074 & 100 \\
        SQL015 & 37376.25 & 40 & 43438 & 50 & 62090 & 10 & 40347.5 & 40 \\
        SQL018 & 1541 & 100 & 825 & 100 & 739.5 & 100 & 1129.5 & 100 \\
        SQL020 & 0 & 100 & 0 & 100 & 231.11 & 100 & 1738 & 100 \\
        \bottomrule
    \end{tabular}
\end{table*}


In the IJON paper, the authors did not specify whether IJON-style annotations can be used to target multiple goals simultaneously.
In our experiments until this point, we had assumed that this was possible and had created annotations that targeted all the vulnerabilities in a project simultaneously.
However, it is possible that IJON-style annotations are only effective when they target a single goal at a time.
To test this hypothesis, we conducted another set of experiments where we created separate copies of a target program with only one vulnerability annotated at a time.

\smallskip
\noindent
\textbf{Dataset.} For this experiment, we selected the \texttt{SQLite3} project from the Magma dataset since it had the largest number of vulnerabilities that were reached by AFL++ (15 in total).
If the annotation-interference hypothesis is correct, this target could potentially exhibit the most pronounced effects due to the high number of vulnerabilities being targeted simultaneously.
To test this hypothesis, we created 15 separate copies of the \texttt{sqlite3\_fuzzer} target program with each copy containing IJON-style annotations for only one of the vulnerabilities.
We then ran 24 hour fuzzing campaigns on each of these copies and repeated it for 10 independent trials.
We repeated this experiment for both manually generated (IJON-SG) annotations and \aijon-generated (\aijon-SG) annotations.

\smallskip
\noindent
\textbf{Result analysis \& findings}
By isolating each vulnerability into its own target program, we expected to see a significant improvement in the energy allocated towards each vulnerability.
This would be reflected in both an increased number of times each vulnerability was reached and potentially a reduced survival time for each vulnerability.
However, our results indicated a more nuanced effect.

We observed that \aijon-SG was able to trigger two vulnerabilities (\texttt{SQL003} and \texttt{SQL013}) that were not detected by \aijon.
Similarly, we observe that IJON-SG was not strictly better than IJON in terms of the number of vulnerabilities triggered or the survival times for these vulnerabilities.
We observed that IJON-SG was able to trigger one vulnerability (\texttt{SQL003}) that was not detected by IJON.
At the same time, it was unable to trigger a vulnerability (\texttt{SQL013}) that was previously detected by IJON.
\aijon-SG on the other hand was able to trigger both of these vulnerabilities.
A detailed comparison of the vulnerabilities triggered by each configuration is provided in Table~\ref{table:single_magma_vuln_comparison}.

We also observed a significant increase in the survival times for \texttt{SQL012}, \texttt{SQL014} and \texttt{SQL015} in \aijon-SG when compared to \aijon.
Two vulnerabilities (\texttt{SQL012} and \texttt{SQL020}) saw an increase in survival time in IJON-SG compared to IJON.
At the same time, \texttt{SQL015} saw a significant reduction in survival time in IJON-SG compared to IJON.
The comparison between the survival times for the vulnerabilities triggered by \aijon, \aijon-SG, IJON and IJON-SG configurations are presented in Table~\ref{table:single_magma_survival_comparison}.

\smallskip
\noindent
\textbf{Analysis metrics.} In order to understand the cause of these results, we utilize the same metrics as described in Section \S\ref{sec:ijon_magma_analysis} to analyze the fuzzing process for each of the vulnerabilities that were triggered by either configuration.
When analyzing the results from the single-goal annotation configuration, we observed that there was an overall increase in the number of times each vulnerability was reached during fuzzing.
In IJON-SG, the biggest increase was seen for \texttt{SQL015} where the average number of times it was reached increased by nearly 100 million compared to IJON which resulted in a significant reduction in the survival time for this vulnerability.

In the case of both \texttt{SQL003} and \texttt{SQL013}, we found that these vulnerabilities were only triggered in less than half of the 10 trials.
This indicates that these vulnerabilities may have been detected or missed purely due to randomness in the fuzzing process rather than the effect of annotations.

In IJON-SG, the two vulnerabilities that had a significantly larger survival time compared to IJON were \texttt{SQL012} and \texttt{SQL020}.
In the case of \texttt{SQL020}, we observed that out of five trials where this vulnerability was triggered, the survival time in four of them was lesser than 100 seconds.
However, there was one outlier trial where the survival time was larger (8350 seconds) which skewed the average survival time for this vulnerability.
This outlier trial may also have been caused due to randomness in the fuzzing process rather than the effect of annotations.

\begin{listing}[h]
\begin{lstlisting}[style=aijonBase, language=C]
IJON_SET(sqlite3VdbeGetOp(v, -1)->opcode); /* PATCHID: 88200 */
MAGMA_LOG("SQL012", sqlite3VdbeGetOp(v,-1)->opcode!=OP_Column);
\end{lstlisting}
\caption{The vulnerable code and IJON annotation for SQL012}%
\label{lst:sql012}
\end{listing}

In the case of \texttt{SQL012}, we observed that the reaching count for this vulnerability was actually lower in IJON-SG compared to IJON.
Upon analyzing the queue of inputs that were saved by the fuzzer after 24 hours of fuzzing, we found that IJON-SG did save multiple inputs that reached the vulnerable location corresponding to \texttt{SQL012} to the corpus.
However, these inputs were not favored by the fuzzer for further mutation evidenced by the fact that we could not find subsequent inputs that were generated directly from mutating these inputs.
This leads us to believe that the annotation was not providing any additional feedback to the fuzzer that would help it distinguish between different inputs that reach this location.

The annotation inserted at this location is shown in Listing~\ref{lst:sql012}.
We manually investigated all the inputs that were saved to the corpus that reached the vulnerable location for \texttt{SQL012} for all 10 trials.
We found that out of 2995 such inputs, 2978 produced the same value for the annotated variable and thus did not provide any additional feedback to the fuzzer.
Since the \texttt{IJON\_SET} annotation only provides new feedback when the value of the annotated variable changes, this annotation can only recognize new inputs if they generate new values for the annotated variable.
Without this, the annotation does not provide any additional feedback to the fuzzer.

\begin{tcolorbox}[colback=gray!10,colframe=black!75,title=Key Observation]
    The annotation \texttt{IJON\_SET} does not provide additional feedback to the fuzzer if the value of the annotated variable does not change significantly across different inputs.
\end{tcolorbox}

\section{Discussion}
Our evaluation of \aijon on the Magma benchmark provides valuable insights into the impact that IJON-style annotations can have on large-scale fuzzing campaigns.
In particular, we have observed a significant reduction in the survival times of certain vulnerabilities when annotations are used.
However, we also observe that this performance improvement is not consistent across all vulnerabilities, and in some cases, the use of annotations can even lead to worse performance compared to the baseline AFL++ configuration without annotations.
We have conducted further experiments to analyze the root cause of this mixed performance and found that the use of annotations can have implicit effects on the fuzzing process in a multi-target setting, which can lead to performance degradation for certain vulnerabilities.

While we have observed that this difference in energy distribution can be mitigated by only annotating a single vulnerability, this approach may not be scalable for large codebases with many target locations that are annotated.
This problem is similar to the problem of Multi-Target Directed Fuzzing which has been studied in the context of directed fuzzing~\cite{huang2024titan}.
A possible future direction may combine a scheduler that recognizes multiple annotated locations and allocates energy to them in a way that mitigates the negative effects of annotations on the fuzzing process.

We have shown that the performance of LLM-generated annotations is comparable to that of manually generated annotations when comparing the survival times of vulnerabilities on the Magma dataset.
This suggests that \aijon can be used as a scalable and automated approach for generating annotations for fuzzing campaigns without significant performance degradation compared to manually generated annotations.
Thus providing a practical solution for further research that studies the impact of annotations on fuzzing campaigns or for practitioners who want to leverage annotations in their fuzzing campaigns without the overhead of manually generating them.

\smallskip
\noindent
\textbf{Threats to Validity.} We discuss the potential threats to the validity of our conclusions by categorizing them into three main categories: external, internal and construct validity.

\textit{External Validity.} Our evaluation uses a subset of target binaries from the Magma benchmark, as we remove any targets without valid seeds or where the fuzzers are unable to generate new inputs.
The subset contains a range of different programs, ranging from database applications to language interpreters to parsers. These applications are in line with typical fuzzing scenarios, but there is a small risk that our observations do not hold for programs from other domains.
Notably, the benchmark's programs contain artifically injected vulnerabilities, which may not reflect the type or distribution of bugs found in software. However, all bugs are based on real-world examples and as such can be considered representative of real-world scenarios.
Our work assumes that a typical security analysis of a large codebase would involve a preliminary static analysis to identify potential vulnerabilities followed by a more intensive analysis to confirm the presence of these vulnerabilities.
We used the \texttt{MAGMA\_LOG} statements as a proxy for the preliminary static analysis to limit the locations where annotations are inserted.
This leads to selecting ideal locations, which static analysis may not achieve in all cases; however, the task of identifying suitable locations or improving static analysis is not the goal of our paper.

\textit{Internal Validity.} To ensure that our findings are not influenced by random factors, we repeated each experiment 10 times and reported the average results.
This mitigates the impact of outliers or random variations in the results and thus allows us to draw more reliable conclusions about the impact of annotations on the fuzzing campaign.
We have implemented the techniques described by Aschermann et al.~\cite{aschermann2020ijon} on top of AFL++ v4.30c, and we used this fuzzer in the same version as baseline for our experiments.
Thus, differences in the results can be attributed to the annotations and not to differences in the underlying fuzzer implementations.

\textit{Construct Validity.} A final threat to validity is whether the evaluation measures what it is intended to measure.
The main objective of our paper is to evaluate the impact of annotations on large-scale fuzzing campaigns.
By analyzing the unique vulnerabilities that are detected as well as their survival times, and by comparing these with a baseline fuzzing campaign without annotations, we can draw conclusions about the impact of annotations on fuzzing campaigns.
While our experiments have used LLMs to generate annotations, we show their impact on the fuzzing campaign is comparable to manually generated annotations.
Still, automated and manual generation of annotations may differ for other targets or scenarios not considered in our work.


\section{Related Work}
Automated vulnerability detection is a vast research area.
In this section we focus on the most relevant prior work in fuzzing, directed fuzzing, the use of annotations in fuzzing and the use of LLMs for code generation in the context of security.

Static analysis have been widely used for detecting potential vulnerabilities in source code as well as in binary executables~\cite{vadayath2022arbiter,gibbs2024operation}

Prior works have attempted to improve fuzzing by using various techniques such as improving seed set selection~\cite{rebert2014optimizing}, mutation strategies~\cite{lyu2019mopt}, coverage weighting~\cite{wang2020not}, limiting the search space for fuzzers~\cite{huang2020pangolin}, using symbolic execution to augment fuzzing~\cite{stephens2016driller,yun2018qsym,poeplau2021symqemu,poeplau2020symbolic} and using grammars to generate target specific inputs~\cite{srivastava2021gramatron}.

Directed fuzzing has also achieved significant success in finding vulnerabilities in specific parts of applications~\cite{bohme2017directed,li2024sdfuzz,luo2023selectfuzz,lee2021constraint,srivastava2022one,liu2024labrador,geretto2025libaflgo}.
Several approaches have also attempted to prioritize targets for directed fuzzing~\cite{weissberg2024sok,rong2024toward,huang2024titan}.

Prior work has also attempted to automatically patch detected crashes to allow fuzzers to detect more bugs occluded by the crash~\cite{raj2024fuzz} as well as using annotations as a bespoke sanitizer for business logic flaws~\cite{wang2025anota}.

Due to the recent advancements in large language models (LLMs), there has been a surge of research that uses LLMs for harness generation~\cite{chen2023hopper,sherman2025no,liu2025promefuzz,lyu2024prompt} as well as for automatically patching vulnerabilities~\cite{yu2025patchagent,nong2025appatch}.
\section{Conclusion}
In this paper, we presented \aijon, a novel and scalable approach for automatically generating IJON-style annotations for fuzzing using large language models.
We evaluate the impact of IJON annotations on large-scale real-world fuzzing campaign by evaluating \aijon on the Magma benchmark suite with the task of reducing the survival time of vulnerabilities.
Our results demonstrate that the performance of \aijon varies significantly across vulnerabilities compared to baseline AFL++.
We conduct further experiments to identify the root cause of the negative impacts that were observed and find that IJON annotations have implicit effects on the fuzzing process in a multi-target setting.
We also observe that LLM generated annotations do perform similar to human generated annotations when comparing the survival times of vulnerabilities on the Magma dataset.
We thoroughly analyze these effects and identify key factors that affect fuzzing performance when using IJON annotations.
\section{Ethical Considerations}
Our work is a replication study that focuses on vulnerability finding. Potential stakeholders include the project maintainers in which bugs are found. 
Our work is based on the MAGMA benchmarks, in which vulnerabilities were artifically injected.  
All the experiments, including those comparing different techniques on the real world vulnerability benchmarks, are conducted locally in isolated environments.
As a result, no new bugs have been found, thus we believe the benefits of studying annotations outweigh any potential risk. 


\section*{Open Science}
All the artifacts including the source code of \aijon, our modified version of Magma are available at \url{https://github.com/bold-rubin/chocolate-milk}.




\cleardoublepage
\bibliographystyle{plain}
\bibliography{strings,biblio}

\cleardoublepage
\appendix

\section{Appendix}

\subsection{Prompt for Annotation Insertion}
In this section, we provide the full prompts that were used for generating annotations.

\begin{listing}[h]
\begin{lstlisting}[style=aijonBase]
You are an experienced C/C++ developer and you have been tasked with adding annotations to the source-code for the \textit{Super Mario Bros.} program as described in the paper IJON: Exploring deep states in fuzzing.
These annotations are expected to guide a fuzzer towards completing one (and only one) level in the game.
A level is completed when Mario reaches the rightmost point in the map.
The fuzzer must recognize unique value in the y coordinates.
You will be provided with the source code of the function.
Read and understand this source code and identify key variables and locations where annotations can help recognize the position of Mario which can in turn help the fuzzer complete one level.

Here is the cheat sheet for ijon annotation
<ijon_cheatsheet>
{{ijon_cheatsheet}}
</ijon_cheatsheet>

here is the example on how to use them

<ijon_example>
{{ijon_example}}
</ijon_example>
\end{lstlisting}
\caption{The prompt used for generating annotations for  \textit{Super Mario Bros.}}%
\label{lst:supermariobros_prompt}
\end{listing}

\begin{listing*}[h]
\begin{lstlisting}[style=aijonBase, numbers=none]
|-------------------------------------------------------------------------------------------------------------------|
|  QUICK REFERENCE (insert exactly as shown with the propery variable(s))                                           |
|-------------------------------------------------------------------------------------------------------------------|
| IJON_CTX((unsigned long long) variable)        - state change occured and all subsequent edges should be          |
|                                                   considered new coverage (use sparingly)                         |
| IJON_INC((int) variable)                       - reward coverage when variable changes                            |
| IJON_SET((int) variable)                       - reward coverage for unique values of variable                    |
| IJON_MAX((unsigned long long) variable)        - reward coverage for maximizing value of variable                 |
| IJON_MIN((unsigned long long) variable)        - reward coverage for minimizing value of variable                 |
| IJON_CMP((unsigned long long) x, (unsigned long long) y)   - reward coverage for integer x satisfying integer     |
|                                                               magic value y (such as header ints)                 |
| IJON_DIST((long long) var_x, (long long) var_y)            - reward coverage for making var_x closer to var_y     |
| IJON_STRDIST((const char *) str_x, (const char *) str_y)   - reward coverage for string str_x getting closer to   |
|                                                               string str_y                                        |
|-------------------------------------------------------------------------------------------------------------------|
\end{lstlisting}
\caption{The IJON cheatsheet given to the LLM for reference}%
\label{lst:ijon_cheatsheet}
\end{listing*}

\begin{listing*}[h]
\begin{lstlisting}[style=aijonBase, language=C]
#include <stdio.h>
#include <string.h>
// Assume IJON runtime is linked and provides the IJON macros described above

int main() {
    // (1) Retrieve some fuzz input values (abstracted for example)
    int x   = get_fuzz_int();         // an integer from fuzz input
    int a   = get_fuzz_int();         // another integer
    int b   = get_fuzz_int();         // another integer
    char *s = get_fuzz_string();      // a fuzzed input string
    
    // (3) IJON_TRACE via INC/SET: expose internal state changes
    static int prev_x = 0;
    if (x != prev_x) {
        IJON_INC(x);                // treat each new value of x as new coverage
        prev_x = x;
    }
    // Now the fuzzer is rewarded for finding inputs that produce new values of x.
    IJON_SET(x);                    // mark this particular value of x as seen (one-time)
    // The fuzzer will try to hit as many unique x values as possible.
    
    // (4) IJON_STATE: incorporate a virtual state for even/odd cases every edge will trigger new coverage for that state after this is set
    if ((x % 2) == 0) {
        IJON_CTX(1);             // enter state "1" for even case
    } else {
        IJON_CTX(2);             // enter state "2" for odd case
    }
    // The following code executes under a state tag (1 or 2) that makes coverage context-sensitive.
    // For example, an inner function might have different behavior in even vs odd state.
    check_complex_condition(a, b); // (some function that may be executed in both contexts)
    // Revert state before exiting the branch (so that states don't leak outside scope)
    if ((x % 2) == 0) IJON_CTX(1); else IJON_CTX(2);
    
    // (5) IJON_CMP: guide bit-by-bit towards a target value
    int key = 0xDEADBEEF;
    IJON_CMP(x, key);             // provide feedback on matching bits between x and 0xDEADBEEF
    IJON_CMP(x, 0xC0DECAFE);      // provide feedback on matching bits between x and 0xC0DECAFE
    if (x == key) {
        printf("Secret unlocked!\\n");
        // ... perhaps trigger a bug here ...
    } else if (x == 0xC0DECAFE) {
        printf("Secret 2 unlocked!\\n");
        // ... perhaps trigger a bug here ...
    }
    
    // (6) IJON_DIST: guide towards satisfying a numeric relation
    IJON_DIST(a + b, 1000);       // reward making (a+b) closer to 1000
    if ((a + b) == 1000) {
        puts("Reached target sum.");
    }
    
    // (7) IJON_STRDIST: guide towards matching a string prefix
    IJON_STRDIST(s, "OPEN");     // reward inputs that match "OPEN" prefix increasingly
    if (strcmp(s, "OPEN") == 0) {
        puts("Opened!");
        // ... maybe a vulnerable condition here ...
    }
    
    // (8) IJON_MAX / IJON_MIN: optimize certain values
    long score = compute_score(s);  
    IJON_MAX(score);            // encourage maximizing the score achieved by input string
    int int_diff = abs(a - b);
    IJON_MIN(int_diff);       // encourage minimizing the difference between a and b
    // With IJON_MAX, the fuzzer will keep inputs that raise 'score'. 
    // With IJON_MIN, it will try to make a and b as close as possible (difference -> 0).
    
    return 0;
}
\end{lstlisting}
\caption{The example program given to the LLM with annotations to use as reference}%
\label{lst:example_program}
\end{listing*}

\begin{listing*}[h]
\begin{lstlisting}[style=aijonBase, numbers=none]
You are a senior fuzz-engineer tasked with assisting a human analyst in instrumenting C/C++ code with IJON macros.
Your goal is to enhance the effectiveness of a coverage-guided fuzzer (such as AFL++) by providing additional, semantically-rich feedback.
This process is based on the technique described in the paper "IJON: Exploring Deep State Spaces via Fuzzing".

Before we begin, here's a cheatsheet of available IJON macros and functions for your reference:

<ijon_cheatsheet>
{{ijon_cheatsheet}}
</ijon_cheatsheet>

# Inputs
You will be provided the code of a C/C++ function along with a diff file that highlights some newly added code which will be referred to as the Point-Of-Interest (POI).
The POI contains some vulnerabilities that are indicated in the MAGMA_LOG statements.
A vulnerability is detected if the conditions indicated in the MAGMA_LOG statements are met.

# Primary objectives
1. Analyze the provided POI and specifically focus on the MAGMA_LOG statements which indicate the condition which leads to the vulnerability.
2. Use IJON annotations to guide the fuzzer towards triggering the location of the vulnerability.
3. Use additional IJON annotations to guide the fuzzer towards triggering the condition which the MAGMA_LOG statements indicate.
4. At the same time, maintain the original semantics of the program.

# Instructions for code instrumentation:
1. Carefully read and understand the provided POI report and the associated C/C++ code of the function.
2. Develop a theory about the vulnerability and the conditions required to trigger it.
3. Insert annotations using IJON macros to guide the fuzzer towards triggering the vulnerability and the conditions leading to it.
4. Avoid adding annotations that are redundant or already covered by edge coverage.
   - For example, each branch of a conditional statement or switch-case is already covered by edge coverage. Adding annotations to each branch is redundant and unnecessary.
5. Do not insert any code that is not directly related to IJON instrumentation.
6. You may not delete or modify lines, only insert.

# Important considerations:
1. Here are some important restrictions and guidelines to follow while instrumenting the code:
   - Do not re-order original side effects.
   - Do not introduce any new code that is not directly related to IJON instrumentation (including conditional statements, function calls, loops).
   - AVOID CALLING FUNCTIONS TO GENERATE THE ARGUMENTS TO ANNOTATE; NO ANNOTATIONS SHOULD BE INVOKING ANY FUNCTION.
   - ANY ANNOTATIONS CONTAINING FUNCTION CALLS MORE THAN JUST THE IJON_XXX CALL WILL BE DELETED AND WASTED. DO NOT DO NOT DO NOT DO NOT DO IT.
      - Exception involve sizeof() which are not function calls but compile-time operators.
   - Do not add or remove any opening or closing braces.
   - Use only standard types (like int, long, size_t, char, etc ) that do not require any headers other than data types that are already used in the code.
   - Keep the original program semantics intact.
   - Do not change the scope of any variables or return statements ESPECIALLY with inline if-statements.
   - Precisely cast values to the expected IJON parameters.
2. Do not insert annotations in between if/else if/else blocks that do not use curly braces or between an if/while/for statement and its curly brace. If a control flow statement has a curly brace and you want to annotate within its block, INSERT AFTER THE CURLY.

Focus on valid variables and try to avoid memory errors in the annotations.
For example, don't insert annotations on struct members by pointer until after the pointer has been checked to be valid in the original code.

Here's an example of how IJON macros can be applied (note that this is over-instrumented for demonstration purposes):

<ijon_example>
{{ijon_example}}
</ijon_example>

When you receive a code block to instrument, follow these steps:
1. Analyze the code thoroughly, considering its overall structure and function.
2. Plan your instrumentation strategy, focusing on the identified vulnerability and key transitions towards triggering it.
3. Insert IJON macros to guide the fuzzer towards the location of the vulnerability as well as towards triggering it.
4. Implement the IJON macros while ensuring you maintain the original program semantics.
5. Format your output using the insertion markers as specified below.

Wrap your instrumentation strategy in <instrumentation_strategy> tags inside your thinking block. In this section:
- List key state transitions, comparisons, and progress indicators you identify in the code.
- Provide a step-by-step plan for applying IJON macros to these identified areas.
- Explain your reasoning for each instrumentation decision.

<insert_output_format>
{{insert_output_format}}
</insert_output_format>

\end{lstlisting}
\caption{The prompt used for generating annotations for Magma benchmark}%
\label{lst:magma_prompt}
\end{listing*}

\subsection{Additional Results}
In this section, we present additional results that were not included in the main paper due to space constraints.
Table~\ref{tab:afl-ijon-reach-number} shows the mean number of inputs until the first bug trigger for AFL and IJON on the Magma benchmark.
This table provides a more detailed comparison of the two fuzzing techniques across different bug IDs, highlighting their performance in terms of reaching bugs.
\begin{table}[t]
    \caption{Mean number of inputs until first bug trigger (AFL vs. IJON vs. Single bug IJON) on Magma.}
    \label{tab:afl-ijon-reach-number}
    \begin{tabular}{lrr}
        \toprule
        \textbf{Bug ID} & \textbf{AFL} & \textbf{IJON} \\
        \midrule
        LUA003 & 1157.80 & 789.00 \\
        LUA004 & 1.00 & 0.80 \\
        PDF003 & 157533.87 & 143248.47 \\
        PDF006 & 2420.57 & 9376.40 \\
        PDF010 & 116572.27 & 160723.23 \\
        PDF011 & 267557482.90 & 315479931.20 \\
        PDF016 & 167722.07 & 210266.70 \\
        PDF018 & 159.90 & 204.10 \\
        PDF019 & 437889.30 & 506384.27 \\
        PDF021 & 10606.23 & 285198.93 \\
        PHP004 & 17685.50 & 57135.90 \\
        PHP009 & 20996.10 & 44755.40 \\
        PHP011 & 2683532.70 & 1396651.60 \\
        PNG001 & 8221735.30 & 27093319.60 \\
        PNG003 & 14155.20 & 20148.60 \\
        PNG007 & 642843.50 & 1233054.00 \\
        SQL002 & 713934.60 & 2137740.00 \\
        SQL003 & N/A & N/A \\
        SQL012 & 7202.30 & 244000.60 \\
        SQL014 & 120326.70 & 164939.30 \\
        SQL015 & 54006728.50 & 80946158.60 \\
        SQL018 & 30921.00 & 10245.00 \\
        SQL020 & 149.40 & 231.40 \\
        SSL002 & 17869.44 & 22464.18 \\
        SSL009 & 69835.18 & 44974.96 \\
        SSL020 & 3347.60 & 3671.30 \\
        TIF002 & 130599.25 & 501558.90 \\
        TIF005 & 0.35 & 0.70 \\
        TIF006 & 0.70 & 0.35 \\
        TIF007 & 613.85 & 662.90 \\
        TIF009 & 46.60 & 62.10 \\
        TIF012 & 202424.60 & 232466.45 \\
        TIF014 & 28655.15 & 25301.30 \\
        XML001 & 821533.20 & 3637.70 \\
        XML003 & 299424.05 & 1176727.65 \\
        XML009 & 301400.65 & 1831588.15 \\
        XML017 & 5084.45 & 4472.40 \\
        \bottomrule
    \end{tabular}
\end{table}

\begin{figure*}[tb!]
    \centering
    \includegraphics[width=\textwidth]{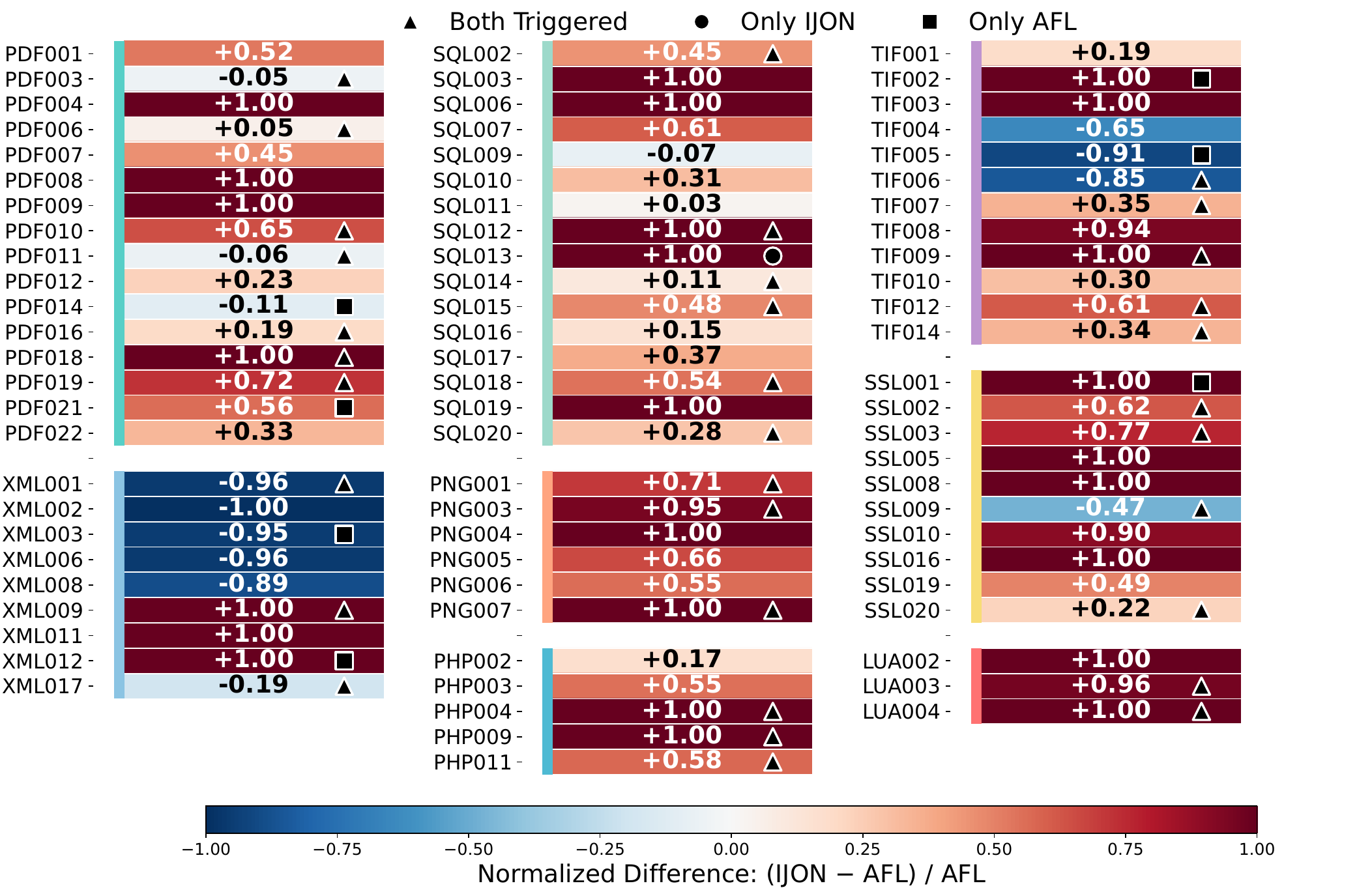}
    \caption{AFL++ vs IJON runtime reach count distribution.}
    \label{fig:ijon_heatmap}
\end{figure*}
\end{document}